# *In Situ* Detection of Active Edge Sites in Single-Layer MoS$_2$ Catalysts

Albert Bruix†, Henrik G. Füchtbauer†, Anders K. Tuxen,
Alex S. Walton, Mie Andersen, Søren Porsgaard,
Flemming Besenbacher, Bjørk Hammer, and Jeppe V. Lauritsen*

Interdisciplinary Nanoscience Center (iNANO), Department of Physics and Astronomy,
*Aarhus University, DK-8000 Aarhus C, Denmark*
*email: jvang@inano.au.dk
† These authors have contributed equally

Graphical abstract:

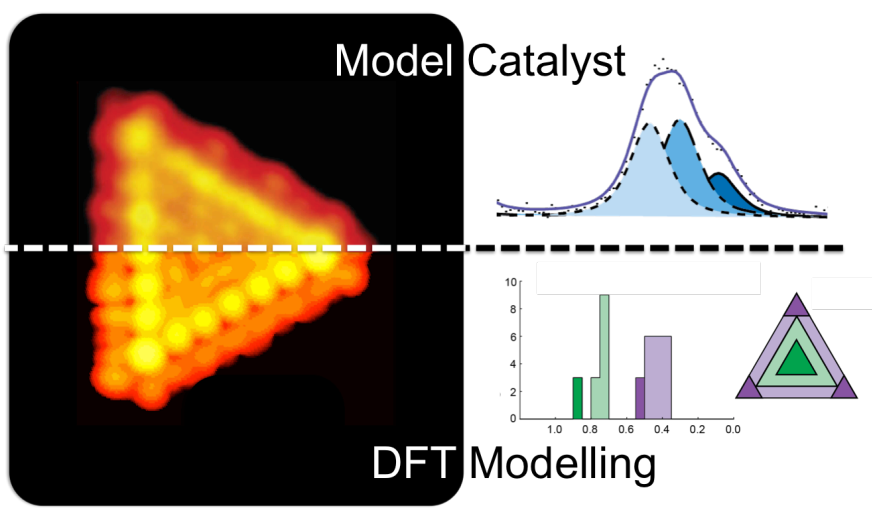

**MoS$_2$ nanoparticles are proven catalysts for processes such as hydrodesulphurization and hydrogen evolution, but unravelling their atomic-scale structure under catalytic working conditions has remained significantly challenging. Ambient pressure X-ray Photoelectron Spectroscopy (AP-XPS) allows us to follow *in-situ* the formation of the catalytically relevant MoS$_2$ edge sites in their active state. The XPS fingerprint is described by independent contributions to the Mo3d core level spectrum whose relative intensity is sensitive to the thermodynamic conditions. Density Functional Theory (DFT) is used to model the triangular MoS$_2$ particles on Au(111) and identify the particular sulphidation state of the edge sites. A consistent picture emerges in which the core level shifts for the edge Mo atoms evolve counter-intuitively towards higher binding energies when the active edges are reduced. The shift is explained by a surprising alteration in the metallic character of the edge sites, which is a distinct spectroscopic signature of the MoS$_2$ edges under working conditions.**

The catalytic chemistry of molybdenite (MoS$_2$) is remarkably versatile. MoS$_2$ has for several decades been in use as a hydrotreating catalyst in oil refineries where it selectively removes both sulphur and nitrogen heteroatoms from the hydrocarbons in crude oil and controls hydrogenation to produce high-value, ultra-clean fuels[1, 2]. More recently, it has been realized that MoS$_2$ also performs well as an inexpensive non-noble metal catalyst for the hydrogen evolution reaction (HER) in (photo)-electrochemical cells for water splitting[3, 4]. The active sites correspond in these cases to the undercoordinated edge sites of the layers in MoS$_2$, which have the capability to dissociate H$_2$ for hydrogenation reactions, to associate H for hydrogen evolution[5], and to bind heteroatoms in hydrocarbon molecules during the hydrogenolysis of S and N. In the industrial hydrotreating catalysts, the activity is optimized by increasing the number of edge sites by dispersing single-layer MoS$_2$ nanoparticles promoted with Co or Ni on a high surface-area γ-alumina support. Stability issues and requirements with respect to conducting properties have led to more elaborate nanostructured MoS$_2$ materials for use in electrochemistry cells with an intrinsically high edge concentration[6] or molecular-like MoS$_2$ clusters consisting of only edge atoms.[7] Fundamental experiments performed on model systems have previously mapped the atomic structure of the edges of single layer MoS$_2$ nanoparticles in high detail using scanning tunnelling microscopy (STM)[8-11]. The STM analysis revealed that the MoS$_2$ edges are significantly reconstructed compared to bulk truncations of MoS$_2$ sheets and that the semiconducting nature of MoS$_2$ is changed to metallic at the edge positions due to electronic edge states.[11, 12] These edge states facilitate the adsorption of various molecules relevant to hydrotreating[13-15] and were also recently shown to be the probable cause of nonlinear optical phenomena in MoS$_2$.[16] Fourier-Transform Infrared Spectroscopy and Temperature Programmed Reduction studies have provided valuable chemical information about these edge sites and were used to follow sulphur reduction and hydrogen uptake[17, 18]. Furthermore, it was possible to elucidate the sulphur coverage of the same edges in industrial-style model catalysts with High-Resolution Transmission Electron Microscopy (HRTEM)[19, 20] again under vacuum conditions. AP-XPS studies performed under electrocatalytic hydrogen evolution conditions recently confirmed the presence of MoS$_2$ as the active catalyst when an amorphous MoS$_3$ precursor was used[21] but a spectroscopic signature of the active MoS$_2$ edges under working conditions is still missing. Here, we use the interplay of AP-XPS, STM and Density Functional Theory (DFT) calculations to elucidate the resulting edge structures which emerge when activating well-characterized single-layer MoS$_2$ nanoparticles with hydrogen gas at elevated pressures and temperatures that resemble hydrotreating catalysis conditions and the reductive environment under hydrogen evolution conditions. We first show that the Mo3d XPS spectra can be described by independent contributions that depend on the position of the Mo atoms within the single-layer MoS$_2$ nanoparticles, where edge Mo atoms exhibit chemical shifts distinct from those in the basal plane. We then identify how such shifts evolve upon reduction of the nanoparticle edges, which enables us to monitor the activation of our MoS$_2$ model catalyst using ambient-pressure XPS. We demonstrate that the fully sulphided starting structure and the active edge structures with a lower S coverage can be discriminated on the basis of XPS data.

**RESULTS AND DISCUSSION**

The large-scale STM image in Figure 1a illustrates the morphology of the MoS$_2$ nanoparticles used in this study. This MoS$_2$ model catalyst system consists of (0001)-oriented single-layer MoS$_2$ nanoparticles (*i.e.* a S-Mo-S layer) on a Au(111) surface and can be synthesized by evaporation of Mo and annealing at 673 K in a 10$^{-6}$ mbar H$_2$S atmosphere. The advantage of using this model system is that it consists of very well-defined MoS$_2$ nanoparticles, the edge structure of which has previously been characterized in atomic detail.[11, 12] The MoS$_2$ nanoparticles have a very strong preference for a triangular shape and are predominantly terminated by the so-called $(10\bar{1}0)$ Mo-edges. The Mo-edges are also established in (S)TEM studies to be by far the most predominant edge type in industrial-type $MoS_2$ catalysts so our model system exposes the catalytically most relevant sites.[19, 20] The $MoS_2$ morphology is in fact slightly sensitive to the size of the nanoparticles[9], but for the coverage of ~0.15 ML illustrated in Figure 1a, approximately 95% of the edge sites expose the $(10\bar{1}0)$ Mo-edges. In the as-synthesized and fully sulphided state seen in the atom-resolved STM image in Figure 1b, the Mo-edge is fully covered with sulphur (100% S) binding $S_2$ dimers at each Mo edge atom (Figure 1c). DFT calculations of realistic computational models such as the one illustrated in Figure 1c allow simulating STM images and obtaining core-level Binding Energy (BE) shifts that can be directly compared to the model catalyst results. These novel representative models simultaneously consider the finite dimension of $MoS_2$ nanoparticles and the presence of the Au(111) support. We have systematically calculated the electronic structure of Au(111)-supported $MoS_2$ nanoparticles of different edge composition representative of the fully sulphided edges (100% S) prepared experimentally (Figure 1a) as well as Mo-edges predicted to be present under $H_2$ reducing conditions[8, 22]. For the as-synthesized $MoS_2$ nanoparticle in Figure 1b, our STM simulations reproduce all characteristic features seen in the experimental image. Such excellent agreement indicates that the theoretical models are fully representative of the experimental $MoS_2$ nanoparticles and can thus be used to perform the interpretation of XPS results.

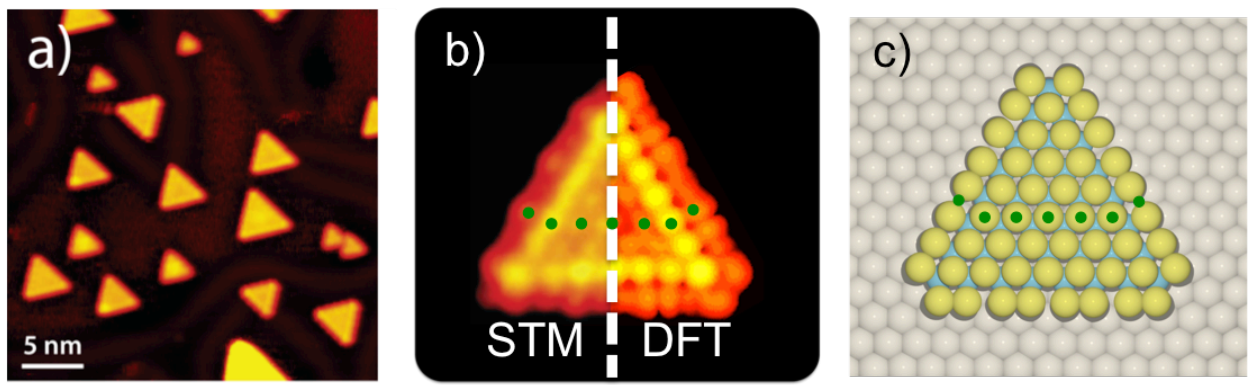

**Figure 1. Model Catalyst and Computational Models:** (a) STM image of $MoS_2$ nanoparticles synthesized on the Au(111) substrate at 0.15 ML coverage and (b) A close-up of a Mo edge terminated $MoS_2$ cluster ($V_{exp}$ = -0.52 V , $I_{exp}$ = 0.44nA) compared to the corresponding STM image simulated by means of DFT and the Tersoff-Hamann approximation ($V_{TH}$ = -0.8 eV, I = 17 pA). The average particle size of $n$ = 8 for a 0.15 ML coverage is reflected, where $n$ is the number of Mo atoms counted along the edge. The STM simulations fully

reproduce all characteristic features seen in the experimental image. This includes the bright metallic brim along the $(10\bar{1}0)$ Mo edges, which is a direct reflection of the particular electronic structure at the Mo edges due to metallic edge state[12], and the off-registry protrusions along the line of basal plane atoms with double periodicity of the S atoms at the edges. The green dots indicate the line of basal plane S atoms and the off-registry protrusion at the edges. The right figure (c) shows the model used for the DFT calculations.

**X-Ray Photoemission Spectroscopy at Ultra-High Vacuum Conditions.**
First we recorded a comparative series of XPS spectra of a cleaved $MoS_2(0001)$ single crystal surface as a reference, industrial-style $MoS_2$ catalyst nanoparticles on a carbon support, the $MoS_2$/Au system and metallic Mo/Au, respectively. The doublet structure of the Mo3d and S2p spectra illustrated in Figure 2 reflects the characteristic spin-orbit splitting between $Mo3d_{3/2}$ and $3d_{5/2}$ states and $S2p_{1/2}$ and $2p_{3/2}$, respectively. The peaks of the reference $MoS_2(0001)$ surface were sharply resolved and symmetric with positions recorded at 229.6 eV for $Mo3d_{5/2}$ and at 162.5 eV for $S2p_{3/2}$, in excellent agreement with literature values for $MoS_2(0001)$[23]. For the $MoS_2$/Au samples the $Mo3d_{5/2}$ main peak lies ~0.4 eV lower in BE at 229.2 eV and the approximately same peak shift is also seen in the S2p signals. This is an indication of partial charge transfer from the substrate, which is also observed in the DFT calculations of $MoS_2$ on Au presented here; upon interaction with the completely sulphided $MoS_2$ nanoparticle, some electron density is depleted from Au and accumulated mainly in the Au-S bonding region. Such charge transfer is expected from XPS experiments from the inverse system (gold on $MoS_2$)[24, 25] and also from the DFT-based studies by Tsai *et al.* involving stripe models of $MoS_2$ on Au(111)[26]. In addition, the Mo3d peaks present a distinct shift and asymmetry towards lower binding energies compared with the $MoS_2(0001)$ reference for the $MoS_2$/Au. A slighter asymmetry is also seen for the $MoS_2$/C technical catalyst system, which also contains nanoparticulate $MoS_2$, but with a larger average size[19]. Furthermore, whereas the S2p doublet is well-resolved on the $MoS_2(0001)$ sample, the $2p^{1/2}$ and $2p^{3/2}$ peaks broaden and overlap in $MoS_2$/C and even more so in $MoS_2$/Au – an indication that S is present in different chemical environments in these samples. In order to elucidate the origin of the asymmetrical peak shape shown in Figure 2, we conducted a high-resolution XPS study for the $MoS_2$/Au system at the Advanced Light Source (Berkeley) using the same synthesis procedure. We start by analyzing the spectra of the as-synthesized sample (Figure 3a), where particles with completely sulphided Mo-edges (100% S) dominate. The peak fitting (see methods) reveals that the asymmetric $Mo3d_{5/2}$ peak is best fitted by three components, *i.e.* two main components at 229.2 eV and 228.8 eV, which are present with a ~1:1 ratio, and a smaller peak at 228.3 eV. The core-level shifts (CLS) calculated with DFT (see methods) are used in order to assign these three peaks, which will henceforth be referred to as high, medium, and low BE components. The calculated Mo3d CLS within the MoS$_2$ nanoparticles with 100%S edges are consistently lower (*i.e.* they are more metallic) for Mo atoms in edge positions than for those in the basal plane (Figure 3b). Such differences divide the calculated spectra into two regions which are separated by ~0.4 eV, in agreement with the separation between the high and medium peak components fitted from the experimental spectra. Therefore, for the as-synthesized MoS$_2$ sample, the high BE component, which is closest to the MoS$_2$(0001) reference, is assigned to basal plane Mo atoms, whereas the medium BE peak is attributed to edge Mo atoms. Previous results are in line with this assignment, where one or more low BE components arise when a bulk MoS$_2$(0001) sample is crushed, leading to larger abundance of undercoordinated edges[23]. However, it should be noted that, as is shown in the forthcoming discussion involving the AP-XPS experiments, these apparently straightforward assignments are subject to variation upon changes in edge composition (sulphidation state) that modify the BE shifts. Importantly, the origin of the 0.4 eV shift to lower BE of the edge Mo atoms can be traced back to their metallic character. This is also directly reflected by the bright brim at the MoS$_2$ edge in the atom-resolved STM image (Figure 1b), which is caused by one-dimensional metallic edge states specific to the 100% Mo-edge. In the projected density of states (pDOS) of a MoS$_2$ nanoparticle, a distinctively higher density across the Fermi level is found for the *d* orbitals of Mo atoms at the edges (Figure S1), but not for those at basal plane positions. The effect resembles the so-called surface core level shift, which is observed in XPS spectra on transition metal surfaces[27] and is a powerful fingerprint for monitoring chemistry on such surfaces [28, 29]. In MoS$_2$, the metallic character of the edges is also directly expressed in the CLS of the Mo atoms, which are thus directly discernable in high-resolution XPS. This enables the use of *in-situ* XPS methods to monitor the nature of the active sites within MoS$_2$-based catalysts. In the S2p spectra of MoS$_2$/Au in Figure 2 and the corresponding high resolution spectrum in Figure 3c we observe a significant broadening compared to the well-resolved S2p doublet of the MoS$_2$(0001) reference, which is indicative of S in several different chemical environments[30]. Our calculated CLS values support this observation, showing a rather wide distribution of S2p shifts within a 0.8 eV range depending on the position of S in the MoS$_2$ particle (Figure 3d). Unlike for amorphous MoS$_3$ and MoS$_2$ [21, 30], which can be distinguished on the basis of their S2p spectra, [21, 30] the scattering of spectroscopically different S species within Au-supported MoS$_2$ nanoparticles is too large to allow a straight-forward interpretation of the spectral shape. Nevertheless, trends are that S in the S$_2$ dimers on the edges are placed 0.1-0.5eV higher in BE relative to the basal plane counterparts. Similarly broad spectra were calculated for the S species at the Mo-edges of MoS$_2$ stripe models, where S2p BEs were found to be larger for the S atoms forming dimers than for monomeric edge S[21]. Another effect in our spectra is that the interaction with the Au support splits the basal plane S atoms into a higher BE component (for the bottom layer S atoms in contact with Au(111)) and a lower BE component (for the upper S layer). The shifts towards larger BE for those S atoms interacting with Au is a result of the formation of S-Au bonds, which drives charge away from Au to the S-Au bonding region and polarises the valence $p_z$ orbitals of the lower S layer towards the Au surface.

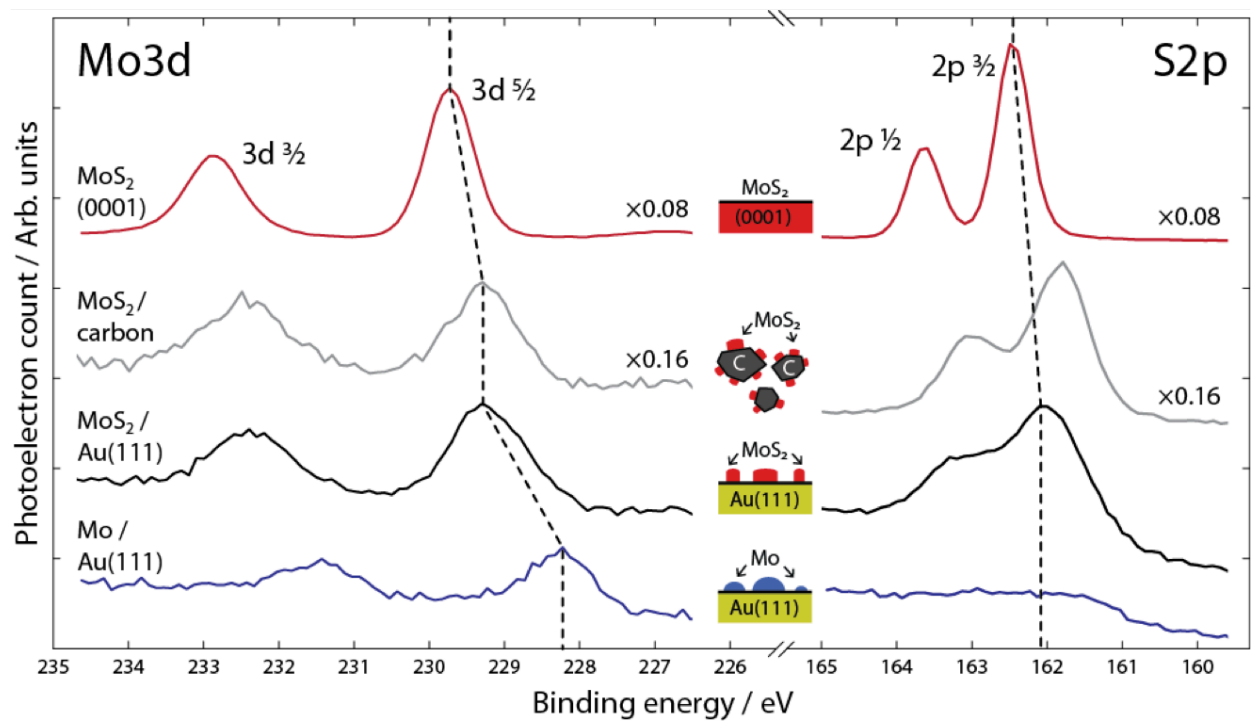

**Figure 2. XPS core level spectra for different MoS$_2$ species:** XPS core level spectra of the Mo3d peaks (3d$_{3/2}$ and 3d$_{5/2}$) and S2p peaks (2p$_{1/2}$ and 2p$_{3/2}$) for a MoS$_2$(0001) surface, MoS$_2$ nanoparticles on a C support, metallic Mo and MoS$_2$ on Au(111), respectively. The spectra have been measured at the SX700 beamline at Aarhus Storage Ring Denmark (ASTRID). Photon energies were hν = 340 eV for Mo3d and hν = 300 eV for S2p. All spectra were recorded in UHV at 298K.

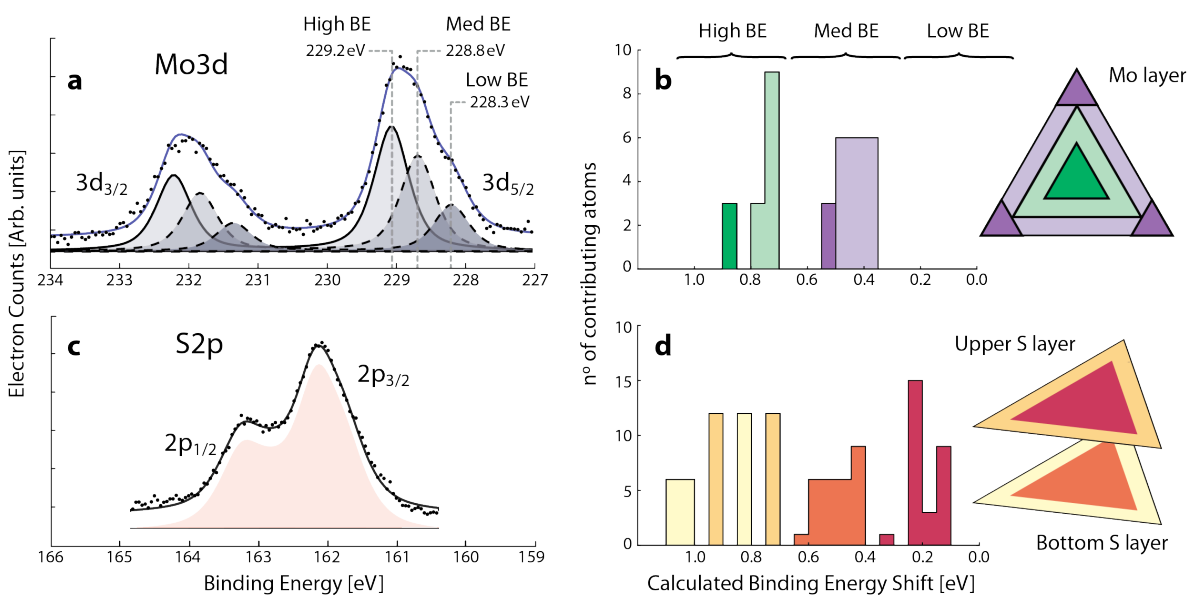

**Figure 3. Deconvolution of the XPS spectra and DFT-calculated shifts:** Mo3d and S2p XPS spectra (a and c) recorded at room temperature and DFT-calculated core-level binding energy (BE) shifts (b and d) for Au(111)-supported $MoS_2$ nanoparticles. The Mo3d fitted spectrum is deconvoluted into three different components (high, medium, and low BE), which are assigned according to the calculated shifts. The different colors within the histograms in the right panels denote the location of Mo and S atoms (corner, edge, near-edge or basal plane) in the $MoS_2$ triangle contributing to the different Mo3d and S2p shifts, according to the color-coding indicated within the corresponding triangle drawings.

**X-Ray Photoemission Spectroscopy at elevated $H_2$ Pressure.**

To follow the activation of the catalyst by exposure to hydrogen gas, we performed AP-XPS on the $MoS_2$/Au sample by sequentially heating it in 0.01 torr $H_2$, recording the Mo3d and S2p XPS spectra for every 50°C from room temperature to 650°C. For these thermodynamic conditions, it is predicted that edge S will be titrated away as $H_2S$ and some H atoms may adsorb on remaining edge S atoms.[8, 18, 31] Therefore, we directly compare the AP-XPS observations with the calculated shifts in BE for the same Au(111)-supported $MoS_2$ nanoparticle models but with different edge compositions. In particular, we have considered Mo-edges with a sequentially lowered S coverage and edges with S-H groups corresponding to 100%S, 50%S, 50%H- 50%S, 25%S, and 0%S (see optimized structures in Figure 5). It should be noted that under UHV conditions (P~$10^{-10}$mbar), STM experiments show that coalescence and decomposition of the $MoS_2$ nanoparticles is observed only for temperatures over 550 °C. In contrast, under the elevated H$_2$ pressures used here the most pronounced changes in the XPS spectra occur in the catalytically interesting temperature range between 200°C and 450°C (Figure 4), where the MoS$_2$ nanoparticle morphology and edge structure in unaffected without the hydrogen gas. The observed changes in the XPS data in Figure 4 do not require any additional peaks for the deconvolution of the Mo3d spectra. Nevertheless, the total intensity of the S2p peak and the ratios between the fitted Mo3d peaks change systematically from the room temperature situation. The Mo3d spectral shape in Figure 4 clearly changes its weight from high to lower BE. In turn, the corresponding S2p spectra (Figure S2) remain broad but the total intensity is gradually reduced, indicating that S is indeed being removed from the MoS$_2$ nanoparticles.

A closer look at the qualitative changes in the Mo3d spectral shapes reveals that exposure to H$_2$ at increasing temperatures initially leads to a decrease in intensity for the medium BE component. This behavior can be explained as a change in edge coverage from 100% S (dimers) towards nanoparticles with 50% S (monomers). The DFT calculations indeed reveal a shift towards higher Mo3d BEs for the Mo atoms at the edges when these are reduced from 100%S to 50%S (Figure 5, red arrow). In the *in-situ* XPS spectra, this appears as a change in the relative intensity of the high and medium BE components. Reduction processes normally lead to the opposite effect in the core level positions, i.e. shifts towards more metallic character, but in this case such an unexpected shift towards higher BE has its origin in the partial loss of the one-dimensional metallic state that is characteristic of fully sulphided $(10\bar{1}0)$ Mo-edges. This is clearly manifested in the simulated STM images, which show that, upon reduction from 100%S to 50%S edge composition, the bright brim along the edge of the MoS$_2$ nanoparticles practically disappears, as the edges become semiconducting (Figure 5). The transition from metallic to non-metallic character of the edges is due to the substitution of the edge $S_2^{\delta-}$ dimers by $S^{2-}$ anions, which have a stronger oxidizing effect on the neighboring Mo atoms. This is also evidenced in the calculated charge density around the different edge S species; the Bader charge of the $S^{2-}$ within the 50%S edge (−0.52e) is significantly more negative than around the S dimers in the 100%S (−0.32e). The described transition is further confirmed by the pDOS on the valence *d* orbitals of the edge Mo atoms (Figure S1). For the 100%S particle, there is a significantly higher pDOS across the Fermi level for Mo atoms at the edges than for those on basal plane positions, whereas for the 50%S particle, the pDOS at the Fermi level is similarly low for Mo atoms in either location. The partial hydrogenation of the 50%S edge which MoS$_2$ particles undergo when exposed to H$_2$ also affects the CLS. The presence of S-H groups mostly perturb the *near*-edge Mo atoms, whose peaks are shifted towards even higher BEs (Figure 5), leading to a situation where the edge Mo are again metallic relative to the basal plane Mo. This is consistent with the presence of metallic edge states for this edge configuration[32] and also reflected by the higher pDOS across the Fermi level (Figure S1).

The AP-XPS Mo3d spectra at further increased temperatures are characterized by a high intensity of the low BE component. At these temperature regimes (beyond 450ºC), highly desulphurized nanoparticle compositions (e.g. 25%S, 0%S) are expected to dominate, which in the DFT calculations give rise to more metallic shifts as the S concentration is reduced (Figure 5, blue arrows). In concordance, these lower sulphided compositions also give rise to higher pDOS on the valence $d$ states across the Fermi level for Mo atoms at the edge positions (Figure S1) and to brighter features in the simulated STM images shown in Figure 5.

The lower graphs in Figure 4 show the quantitative area changes of the Mo3d peak fit components and the S2p spectral intensity as a function of temperature. The main observation for the Mo3d spectra is that the medium BE (circles) and high BE (triangles) components decrease in intensity, while the intensity of the low energy component (squares) grows. The initial decrease of the two higher energy peaks is slow, but from 200°C and onwards up to 450°C the slope of the middle BE becomes much steeper and this component drops by almost 50%. At the same time, the low BE signal increases, indicating that some of the particles are reduced to less sulphided states, in particular above 300°C. Overall, this is consistent with the conversion of the middle BE component associated with Mo-edges with 100% S into edges with 50%S coverage or lower within the catalytically interesting temperature interval. We do not consider a sintering process to be the likely explanation for the changes in basal plane to edge ratio, since STM investigations have shown that the initial size of the $MoS_2$ triangles are unaffected by extensive hydrogen exposure (hours) at $10^{-6}$ mbar and 450ºC. It should be noted, that the overall process also results in a drop of the total Mo3d spectral area, which likely reflects a loss of Mo due to accelerated alloying between the gold substrate and the molybdenum atoms exposed by the reducing conditions[33]. As a result of the intense desulphurization which occurred in the experiment without a balancing source of S (e.g. S-containing hydrocarbons as in hydrotreating), the $MoS_2$ nanoparticles may even have decomposed beyond the reduced edges seen in Figure 5 at the highest temperatures (>500ºC). In fact, the S2p/Mo3d ratio decreases by approximately a factor 2.2 from the first to the last measurement in the series. This change cannot be explained by removal of only edge S in a representative $n = 8$ $MoS_2$ triangle (Figure 1) and at high temperatures in our experiment the reaction must have proceeded further than this, thereby desulphurizing the near-edge region of the basal plane.

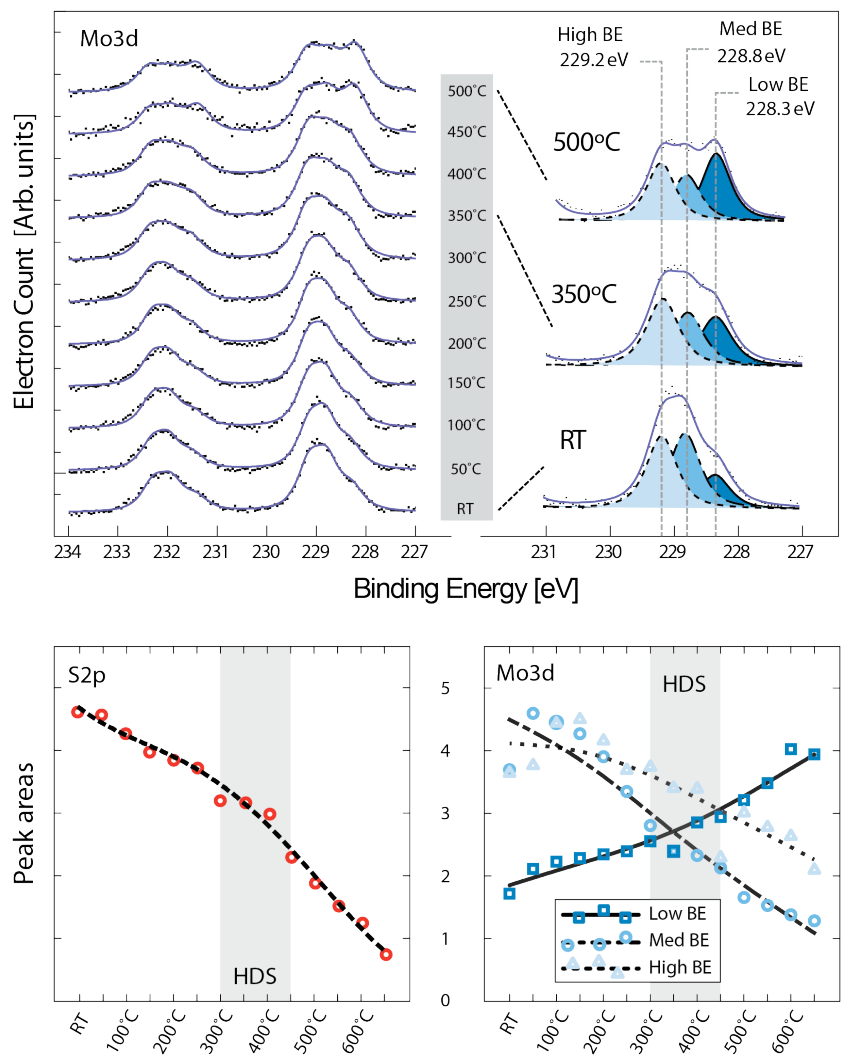

**Figure 4.** *In-situ* monitoring of the edge configuration through AP-XPS spectra: Peak fits of the Mo3d spectra between RT and 500°C (upper left), together with plot of the Mo3d and S2p peak areas determined from the peak fits (bottom), in the range from RT to 650°C. The upper right panel shows a close-up of the devoncoluted Mo3d$_{5/2}$ spectra for three characteristic temperatures (Room Temperature, 350ºC, and 500ºC). These plots close-ups clearly show the differences in the relative intensities of the medium and high BE shifts, which are due to the evolution towards the high BE region experienced by the edge Mo atoms upon reduction from 100%S edge composition to 50%S. The shaded area denoted 'HDS' indicates the typical temperatures used in industrial hydrotreating catalysis.

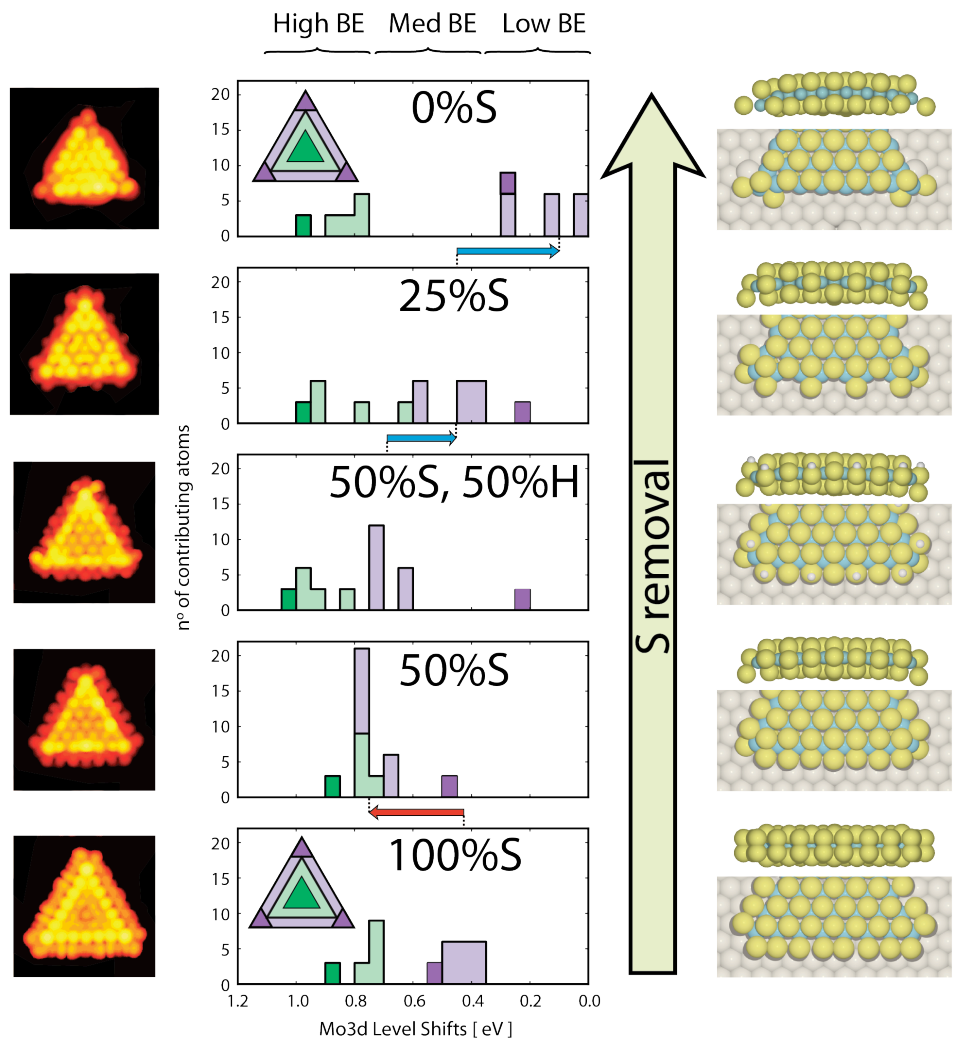

**Figure 5.** Simulated STM spectra and core-level shifts: Evolution of the calculated Mo3d binding energy shifts for different positions within $n = 8$ MoS$_2$ nanoparticles of varying edge composition. From bottom to top, models with edge compositions of 100%S, 50%S, 50%S-50%H, 25%S, and 0%S have been considered. The triangles within the top and bottom graphs are color-coded according to the corresponding Mo3d BE shifts. To the left are shown STM images calculated with the Tersoff-Hamann approach at I = 17 pA and V = −0.8 eV. The red and blue arrows indicate the direction of the changes in BE shift for the edge sites during S removal.

CONCLUSIONS

We have shown that the detailed spectral shape of Mo3d core level results from the nanoscale structure of $MoS_2$. Moreover, high-resolution XPS shows that the spectral shape is sensitive to the S coordination at the edges, and this effect enables characterisation of the transitions involved in desorption of S and adsorption of H that describes the catalytically active $MoS_2$ edges. Remarkably, the edge reduction process from 100% to 50% sulphur coverage leads to a shift towards higher Mo3d BE, which is opposite to the conventional interpretation of 'oxidation' in CLS. The shift can be traced back to variations in the metallic character of the Mo edge, which is lost when $S_2^{\delta-}$ dimers are replaced by more reducing $S^{2-}$ species. We anticipate that the chemical shifts seen in the molybdenum core levels by XPS should also be observable in Near-edge X-ray absorption (NEXAFS) or Resonant Inelastic X-ray Scattering (RIXS) experiments, which are already powerful tools for the characterization of catalysts in the working state[34]. These techniques also probe the electron levels near the Fermi level, and should therefore be directly sensitive to the changes in the metallic character of the edges during formation of the active sites. They are furthermore applicable for characterization of technical catalysts beyond the planar model catalysts approach applied in the present studies, and since the observed change in the metallic character is predicted to be independent of the Au support[32] the present work thus provides a general framework for interpretation of *in-situ* x-ray spectroscopy of $MoS_2$ edges.

## METHODS

**$MoS_2$/Au(111) model system:** The STM experiments were performed in an ultra-high vacuum (UHV) chamber equipped with a home built Aarhus STM. The XPS measurements were measured at two different beamlines, the SX700 beamline at ASTRID (Aarhus Storage Ring, Denmark) and the 11.0.2 beamline at the ALS (Advanced Light Source, Lawrence Berkeley National Laboratory). For direct comparison and to avoid contamination problems due to air exposure, the $MoS_2$ nanoparticles were synthesized *in-situ* for each experiment in either of the UHV chambers using a pre-calibrated Mo e-beam evaporator (Oxford Research EGN4 or EGCO4), a homemade $H_2S$ doser[35] on the same freshly prepared Au(111) sample. The synthesis is performed in two steps, first by e-beam evaporation of Mo in a sulphiding atmosphere ($H_2S$, $P=10^{-6}$ mbar) followed by continued sulphidation at elevated temperature (450°C) (see details in [9, 11]). We synthesized ~0.15 ML of $MoS_2$ on the gold substrate, where 1 ML is equivalent to a fully covered substrate. Regular checks were made for drift in the Mo flux with the flux reading of the e-beam evaporator and a quartz crystal microbalance (ALS).

**X-ray Photoelectron Spectroscopy:** We have recorded spectra of the Mo3d peak with a photon energy of 370 eV and the S2p peak with 300 eV. For each of the Mo3d and S2p spectra a Au4f spectrum was recorded with the same photon energy to allow energy calibration. All XPS peaks were fitted with pseudo-Voigt line-shapes, consisting of a sum of a Gaussian and Lorentzian contribution, using the same method as implemented by others [36]. For the $MoS_2$ nanoparticles we use symmetric peaks while asymmetric peaks were used for metallic molybdenum, consistent with the Fermi sea screening of the core hole due to the coupling to the gold substrate's free electrons[37]. A linear background was subtracted from all spectra. The splitting between the Mo $3d_{3/2}$ and $3d_{5/2}$ peaks was determined to be 3.16 eV. This is consistent with the value reported in literature.[23, 38]. For each spin-orbit doublet the ratio between the two peaks was bound narrowly around the ratio given from quantum numbers, e.g. 2:3 for the Mo3d peak. The Au4f spectra were fitted with two asymmetric components, where the highest binding energy corresponding to the bulk component was assigned a binding energy of 84.0 eV to which the intensity and energy of the rest of the spectra were calibrated.

**DFT Calculations**: The DFT calculations were performed using the PBE exchange-correlation functional, the Projector Augmented Wave (PAW) method [39], and a real space grid for the expansion of the wave functions, as implemented in the GPAW code [40]. In order to calculate the relative binding energy of the Mo3d and S2p electrons for different Mo and S atomic species, a PAW-type pseudopotential with an electron hole in the selected level was used for describing the atom the core electron was removed from. A compensating homogeneous background density was included in order to compensate the positive charge resulting from the removal of an electron from a core level. Prior to the generation of the core-holes, the geometries for the different models used were optimised until the forces on each atom were lower than 0.025 eV/Å. The upper layer of the two-layer slab used to represent the Au(111) substrate was also allowed to relax. For each optimised system, the generation of an electron hole was considered for all nonequivalent atomic positions. The chemical shifts of Mo3d were calculated relative to the average shift of Mo atoms within an Au(111)-supported $Mo_{21}$ cluster, which corresponds to metallic Mo. A spectator Mo atom located at the opposite side of the slab of each system was used to calculate the relative shifts between each $MoS_x$ nanostructure and the metallic $Mo_{21}$ cluster. The shifts of S2p were calculated relative to a spectator S atom on the opposite side of the slab. The STM simulations in Figures 1 and 5 are based on the Tersoff-Hamann approximation, which relates the intensities of STM images with isosurfaces of the local density of states at the Fermi level[41, 42].

**Acknowledgements**

We thank Jess Staussholm-Møller for initiation and assistance with the DFT calculations in their early stages. Michael Brorson, Poul Georg Moses and Stig Helveg are acknowledged for discussions and for providing the industrial-style $MoS_2$/carbon samples. We acknowledge support from Haldor Topsøe A/S, the European Research Council (ERC Grant no. 239834 and The Danish Council for Strategic Research (CAT-C) and the Danish Council for Independent Research | Natural Sciences. AB acknowledges support from the European Research Council under the European Union's Seventh Framework Programme (FP/2007-2013) / Marie Curie Actions / Grant no. 626764 (Nano-DeSign). We acknowledge beamtime received on the SX-700 beamline at ASTRID (Aarhus Denmark) and the environmental science beam line 11.0.3 at the Advanced Light Source (ALS) at Berkeley Lab. The Advanced Light Source is supported by the Director, Office of Science, Office of Basic Energy Sciences, of the U.S. Department of Energy under Contract No. DE-AC02-05CH11231. We thank Z. Li and H. Bluhm for helpfulness and technical assistance at ASTRID and ALS, respectively.

**Competing Financial Interests**

The authors declare no competing financial Interests.

**References**

1. Topsøe, H.; Clausen, B. S.; Massoth, F. E. *Hydrotreating Catalysis*. Springer: 1996.
2. Toulhoat, H.; Raybaud, P. *Catalysis by Transition Metal Sulphides*. Editions Technip, Paris: 2013.
3. Jaramillo, T. F.; Jørgensen, K. P.; Bonde, J.; Nielsen, J. H.; Horch, S.; Chorkendorff, I. Identification of Active Edge Sites for Electrochemical $H_2$ Evolution from $MoS_2$ Nanocatalysts. *Science* **2007**, 317, 100-102.
4. Kibsgaard, J.; Chen, Z. B.; Reinecke, B. N.; Jaramillo, T. F. Engineering the Surface Structure of $MoS_2$ to Preferentially Expose Active Edge Sites for Electrocatalysis. *Nat. Mater* **2012**, 11, 963-969.
5. Hinnemann, B.; Moses, P. G.; Bonde, J.; Jørgensen, K. P.; Nielsen, J. H.; Horch, S.; Chorkendorff, I.; Nørskov, J. K. Biomimetic Hydrogen Evolution: $MoS_2$ Nanoparticles as Catalyst for Hydrogen Evolution. *J. Am. Chem. Soc.* **2005**, 127, 5308-5309.
6. Kibsgaard, J.; Jaramillo, T. F.; Besenbacher, F. Building an Appropriate Active-Site Motif into a Hydrogen-Evolution Catalyst with Thiomolybdate $Mo_3S_{13}(^{2-})$ Clusters. *Nat. Chem.* **2014**, 6, 248-253.
7. Karunadasa, H. I.; Montalvo, E.; Sun, Y.; Majda, M.; Long, J. R.; Chang, C. J. A Molecular $MoS_2$ Edge Site Mimic for Catalytic Hydrogen Generation. *Science* **2012**, 335, 698-702.
8. Lauritsen, J. V.; Bollinger, M. V.; Lægsgaard, E.; Jacobsen, K. W.; Nørskov, J. K.; Clausen, B. S.; Topsøe, H.; Besenbacher , F. Atomic-Scale Insight into Structure and Morphology Changes of $MoS_2$ Nanoclusters in Hydrotreating Catalysts. *J. Catal.* **2004**, 221, 510-522.
9. Lauritsen, J. V.; Kibsgaard, J.; Helveg, S.; Topsøe, H.; Clausen, B. S.; Besenbacher , F. Size-Dependent Structure of $MoS_2$ Nanocrystals. *Nat. Nanotechnol.* **2007**, 2, 53-58.
10. Tuxen, A.; Kibsgaard, J.; Gøbel, H.; Lægsgaard, E.; Topsøe, H.; Lauritsen, J. V.; Besenbacher, F. Size Threshold in the Dibenzothiophene Adsorption on $MoS_2$ Nanoclusters. *ACS Nano* **2010**, 4, 4677-4682.
11. Helveg, S.; Lauritsen, J. V.; Lægsgaard, E.; Stensgaard, I.; Nørskov, J. K.; Clausen, B. S.; Topsøe, H.; Besenbacher, F. Atomic-Scale Structure of Single-Layer $MoS_2$ Nanoclusters. *Phys. Rev. Lett.* **2000**, 84, 951-954.
12. Bollinger, M. V.; Lauritsen, J. V.; Jacobsen, K. W.; Nørskov, J. K.; Helveg, S.; Besenbacher, F. One-Dimensional Metallic Edge States in $MoS_2$. *Phys. Rev. Lett.* **2001**, 87, 196803.
13. Lauritsen, J. V.; Nyberg, M.; Vang, R. T.; Bollinger, M. V.; Clausen, B. S.; Topsøe, H.; Jacobsen, K. W.; Besenbacher, F.; Lægsgaard, E.; Nørskov, J. K.*, et al.* The Chemistry of One-Dimensional Metallic Edge States in $MoS_2$ Nanoclusters. *Nanotech.* **2003**, 14, 385-389.
14. Temel, B.; Tuxen, A. K.; Kibsgaard, J.; Topsøe, N. Y.; Hinnemann, B.; Knudsen, K. G.; Topsøe, H.; Lauritsen, J. V.; Besenbacher, F. Atomic-Scale Insight into the Origin of Pyridine Inhibition of $MoS_2$-Based Hydrotreating Catalysts. *J. Catal.* **2010**, 271, 280-289.
15. Tuxen, A. K.; Fuchtbauer, H. G.; Temel, B.; Hinnemann, B.; Topsøe, H.; Knudsen, K. G.; Besenbacher, F.; Lauritsen, J. V. Atomic-Scale Insight into Adsorption of Sterically Hindered Dibenzothiophenes on $MoS_2$ and Co-Mo-S Hydrotreating Catalysts. *J. Catal.* **2012**, 295, 146-154.
16. Yin, X. B.; Ye, Z. L.; Chenet, D. A.; Ye, Y.; O'Brien, K.; Hone, J. C.; Zhang, X. Edge Nonlinear Optics on a $MoS_2$ Atomic Monolayer. *Science* **2014**, 344, 488-490.
17. Topsøe, N.-Y.; Topsøe, H. Ftir Studies of $Mo/Al_2O_3$-Based Catalysts: 2. Evidence for the Presence of S-H Groups and Their Role in Acidity and Activity. *J. Catal.* **1993**, 139, 641-651.
18. Dinter, N.; Rusanen, M.; Raybaud, P.; Kasztelan, S.; da Silva, P.; Toulhoat, H. Temperature-Programed Reduction of Unpromoted $MoS_2$-Based Hydrodesulfurization Catalysts: Experiments and Kinetic Modeling from First Principles. *J. Catal.* **2009**, 267, 67-77.
19. Hansen, L. P.; Ramasse, Q. M.; Kisielowski, C.; Brorson, M.; Johnson, E.; Topsøe, H.; Helveg, S. Atomic-Scale Edge Structures on Industrial-Style $MoS_2$ Nanocatalysts. *Angew. Chem. Int. Ed.* **2011**, 50, 10153-10156.

20. Kisielowski, C.; Ramasse, Q. M.; Hansen, L. P.; Brorson, M.; Carlsson, A.; Molenbroek, A. M.; Topsøe, H.; Helveg, S. Imaging $MoS_2$ Nanocatalysts with Single-Atom Sensitivity. *Angew. Chem. Int. Ed.* **2010**, 49, 2708-2710.
21. Casalongue, H. G. S.; Benck, J. D.; Tsai, C.; Karlsson, R. K. B.; Kaya, S.; Ng, M. L.; Pettersson, L. G. M.; Abild-Pedersen, F.; Norskov, J. K.; Ogasawara, H*., et al.* Operando Characterization of an Amorphous Molybdenum Sulfide Nanoparticle Catalyst During the Hydrogen Evolution Reaction. *J. Phys. Chem. C* **2014**, 118, 29252-29259.
22. Raybaud, P.; Hafner, J.; Kresse, G.; Kasztelan, S.; Toulhoat, H. *Ab Initio* Study of the $H_2$-$H_2S/MoS_2$ Gas-Solid Interface: The Nature of the Catalytically Active Sites. *J. Catal.* **2000**, 189, 129-146.
23. Mattila, S.; Leiro, J. A.; Heinonen, M.; Laiho, T. Core Level Spectroscopy of $MoS_2$. *Surf. Sci.* **2006**, 600, 5168-5175.
24. Lince, J. R.; Carre, D. J.; Fleischauer, P. D. Schottky-Barrier Formation on a Covalent Semiconductor without Fermi-Level Pinning - the Metal $MoS_2(0001)$ Interface. *Phys. Rev. B* **1987**, 36, 1647-1656.
25. Shi, Y.; Huang, J.-K.; Jin, L.; Hsu, Y.-T.; Yu, S. F.; Li, L.-J.; Yang, H. Y. Selective Decoration of Au Nanoparticles on Monolayer $MoS_2$ Single Crystals. *Sci. Rep.* **2013**, 3.
26. Tsai, C.; Abild-Pedersen, F.; Nørskov, J. K. Tuning the $MoS_2$ Edge-Site Activity for Hydrogen Evolution Via Support Interactions. *Nano Lett.* **2014**, 14, 1381-1387.
27. Spanjaard, D.; Guillot, C.; Desjonquères, M.-C.; Tréglia, G.; Lecante, J. Surface Core Level Spectroscopy of Transition Metals: A New Tool for the Determination of Their Surface Structure. *Surf. Sci. Rep.* **1985**, 5, 1-85.
28. Baraldi, A. Structure and Chemical Reactivity of Transition Metal Surfaces as Probed by Synchrotron Radiation Core Level Photoelectron Spectroscopy. *J. Phys. Cond. Matt.* **2008**, 20, 093001.
29. Gustafson, J.; Borg, M.; Mikkelsen, A.; Gorovikov, S.; Lundgren, E.; Andersen, J. N. Identification of Step Atoms by High Resolution Core Level Spectroscopy. *Phys. Rev. Lett.* **2003**, 91, 056102.
30. Weber, T.; Muijsers, J. C.; Niemantsverdriet, J. W. Structure of Amorphous $MoS_3$. *J. Phys. Chem.* **1995**, 99, 9194-9200.
31. Schweiger, H.; Raybaud, P.; Kresse, G.; Toulhoat, H. Shape and Edge Sites Modifications of $MoS_2$ Catalytic Nanoparitcles Induced by Working Conditions: A Theoretical Study. *J. Catal.* **2002**, 207, 76-87.
32. Bollinger, M. V.; Jacobsen, K. W.; Nørskov, J. K. Atomic and Electronic Structure of $MoS_2$ Nanoparticles. *Phys. Rev. B* **2003**, 67, 085410.
33. Potapenko, D. V.; Horn, J. M.; Beuhler, R. J.; Song, Z.; White, M. G. Reactivity Studies with Gold-Supported Molybdenum Nanoparticles. *Surface Science* **2005**, 574, 244-258.
34. Weckhuysen, B. M. *In-Situ Spectroscopy of Catalysts*. American Scientific Publishers: 2004.
35. Lauritsen, J. V.; Besenbacher , F. Model Catalyst Surfaces Investigated by Scanning Tunneling Microscopy. *Adv Catal* **2006**, 50, 97-147.
36. Fairley, N.; Carrick, A. *The Casa Cookbook - Part 1: Recipes for Xps Data Processing*. Acolyte Science: 2005; p 368.
37. Hüfner, S.; Wertheim, G. K.; Wernick, J. H. Xps Core Line Asymmetries in Metals. *Solid State Commun* **1975**, 17, 417-422.
38. Wagner, C. D.; Riggs, W. M.; Davis, L. E.; Moulder, J. F.; Muilenberg, G. E. *Handbook of X-Ray Photoelectron Spectroscopy*. Perkin-Elmer Corp, Edina, Minn: 1979.
39. Blöchl, P. E. Projetor Augmented Wave Method. *Phys. Rev. B* **1994**, 50, 17953-17979.
40. Enkovaara, J.; Rostgaard, C.; Mortensen, J. J.; Chen, J.; Dułak, M.; Ferrighi, L.; Gavnholt, J.; Glinsvad, C.; Haikola, V.; Hansen, H. A*., et al.* Electronic Structure Calculations with Gpaw: A Real-Space Implementation of the Projector Augmented-Wave Method. *J. Phys. Cond. Matt.* **2010**, 22, 253202.

41. Tersoff, J.; Hamann, D. R. Theory and Application for the Scanning Tunneling Microscope. *Phys. Rev. Lett.* **1983**, 50, 1998-2001.
42. Tersoff, J.; Hamann, D. R. Theory of the Scanning Tunneling Microscope. *Phys. Rev. B* **1985**, 31, 805-813.

# Supplementary Information:
# *In-situ* Detection of Active Edge Sites
# in Single-Layer MoS$_2$ Catalysts

Albert Bruix†, Henrik G. Füchtbauer†, Anders K. Tuxen,
Alex S. Walton, Mie Andersen, Søren Porsgaard,
Flemming Besenbacher, Bjørk Hammer, and Jeppe V. Lauritsen*

*Interdisciplinary Nanoscience Center (iNANO), Department of Physics and Astronomy,*
*Aarhus University, DK-8000 Aarhus C, Denmark*
*email: jvang@inano.au.dk*
*† These authors have contributed equally*

The supplementary material includes Figures S1 and S2.

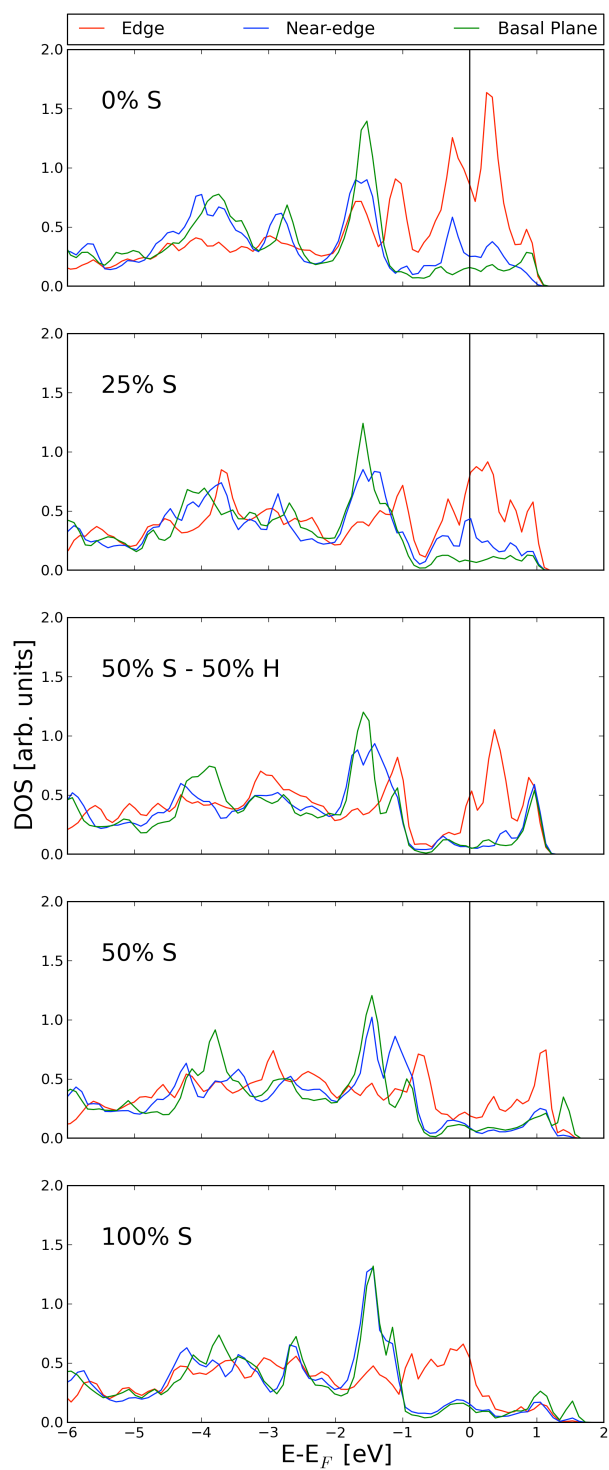

**Figure S1 | Evolution of the metallic character of Mo atoms at different sites**: DOS projected on the *d* states of Mo atoms at edge (red), near-edge (blue), and basal plane (green) positions for different $MoS_2$ nanoparticle compositions.

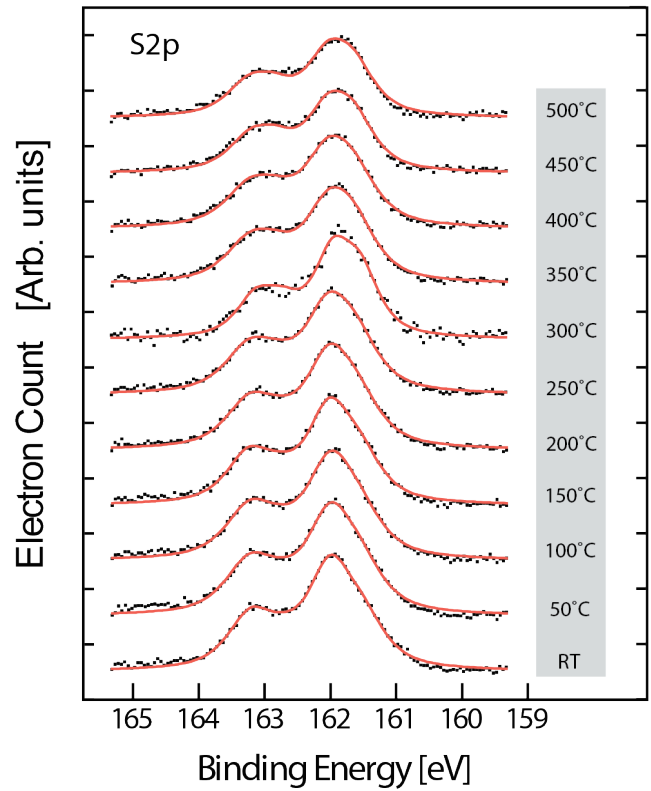

**Figure S2 | *In-situ* monitoring of the edge configuration through AP-XPS spectra:** Peak fits of the S2p spectra between RT and 500°C. The corresponding S2p peak areas determined from the peak fits in the range from RT to 650°C are shown in Figure 4 (bottom left).